\documentclass[a4paper,11pt]{article}
\usepackage{pos}

\RequirePackage{color}
\definecolor{dkgreen}{rgb}{0.2,0.7,0.4}
\definecolor{dkblue}{rgb}{0.2,0.2,0.7}
\definecolor{dkred}{rgb}{0.8,0,0}
\definecolor{dkgreen}{rgb}{0.2,0.8,0.4}

\title{Meron-Cluster Algorithms for Quantum Link Models}

\author*[a]{Joao C. Pinto Barros}
\author[a]{Thea Budde}
\author[a]{Marina Krstic Marinkovic}

\affiliation[a]{Institut f\"{u}r Theoretische Physik, Wolfgang-Pauli-Straße 27, ETH Z\"{u}rich, 8093 Z\"{u}rich, Switzerland}

\emailAdd{jpinto@phys.ethz.ch}
\emailAdd{tbudde@phys.ethz.ch}
\emailAdd{marinama@ethz.ch}

\abstract{
State-of-the-art algorithms for simulating fermions coupled to gauge fields often rely on integrating fermion degrees of freedom. While successful in simulating QCD at zero chemical potential, at finite density these approaches are hindered by the sign problem. 
We discuss the simulation of lattice gauge theories in the Hamiltonian formalism and present a generalized meron-cluster algorithm for the simulation of the $U\left(1\right)$ Quantum Link Model for spin $1/2$. This enables the study of models directly relevant to current quantum simulators and is a promising first step toward constructing new efficient algorithms for more complicated gauge theories.
}

\FullConference{%
  The 40th International Symposium on Lattice Field Theory (Lattice 2023)\\
  July 31st - August 4th, 2023\\
  Fermi National Accelerator Laboratory
}

\begin{document}
\maketitle

\section{Introduction}

The Hybrid Monte Carlo algorithm has been one of the instrumental developments that have allowed for the current success of 
lattice Quantum Chromodynamics (QCD), and can be applied to other lattice field theories~\cite{krieg2018pqh,buividovich2018hybrid,del2016large,bennett2022lattice,bennett2018sp}. The process involves the integration of fermions, giving rise to an effective bosonic model. The introduction of a finite chemical potential or a topological $\theta$ term leads to a sign problem and constitutes a limitation of the current methods \cite{duane1987hybrid,gattringer2009quantum}. 

Quantum simulators hold the promise to surpass these problems by mapping these theories directly to quantum variables. In standard approaches, the theory is formulated in the Hamiltonian formulation. This Hamiltonian can then be simulated using appropriate degrees of freedom in digital or analog quantum simulators, see e.g. \cite{banuls2020simulating}. 

The Hamiltonian formulation offers an opportunity to construct suitable quantum simulators and provides an alternative formulation where classical algorithms can be developed. While historically not the preferred route for Monte Carlo simulations in particle physics, these types of simulations are wildly abundant in condensed matter theory, for example \emph{Quantum Monte Carlo}~(QMC) approaches. The fermionic sign problems can also occur in QMC and they constitute the same type of obstacles faced in the Lagrangian formulation. Notably, there is a class of models where the sign problem can be solved under QMC using the \emph{meron-cluster algorithms} \cite{chandrasekharan1999meron,chandrasekharan2000meron,chandrasekharan2000critical,cox2000meron,osborn2001superconductivity,chandrasekharan2002kosterlitz,liu2021hamiltonian}. Presumably, formulating QCD in the Hamiltonian formulation can offer new approaches that surpass standard methods. In practice, we still lack a way to address this type of questions.

The Schwinger model shares several features with Quantum Chromodynamics including confinement and the existence of a topological $\theta$ term. Due to its simplicity, it is a popular model used for the development of novel classical methods and a common target of quantum simulations. For the latter, it is particularly important to formulate the theory in a finite Hilbert space. Quantum Link Models provide such formulation without breaking gauge invariance \cite{wiese2022quantum}. In the simplest example, the $U\left(1\right)$ gauge variables that live on the links are represented by $1/2$ quantum spins, hosting a local 2-dimensional Hilbert space.  In this work, we develop an adaptation of the meron-cluster algorithm, that allows for the simulation of this model. Pioneering works in this direction have relied on the emergence of Gauss' law in an appropriate parameter regime, rather than a direct implementation \cite{frank2020emergence,banerjee2023quantum,huffman2023cluster}. A key feature is the ability to satisfy Gauss’ law exactly during cluster flips, allowing for non-local updates while remaining within the physical subspace. These results are both relevant to classical and quantum simulations of gauge theories, and enable explorations of models directly relevant to current state-of-the-art quantum simulators. 


\section{Spin-$1/2$ Quantum Link Models in $1+1$ Dimensions}
\label{sec:QLM}

The spin-$1/2$ quantum link model, with staggered fermions, in $1+1$ dimensions is given by 

\begin{equation}
    \label{eq:H}
    H=-t\sum_{n=0}^{L-1}
    (c^\dagger_n S^+_n c_{n+1}+\mathrm{h.c.})
    +m\sum_{n=0}^{L-1}\left(-1\right)^nc^\dagger_nc_{n}
    +U\sum_{n=0}^{L-1}
    \left(c^\dagger_nc_n-\frac{1}{2}\right)
    \left(c^\dagger_{n+1}c_{n+1}-\frac{1}{2}\right).
\end{equation}
The operators $c_n$/$c^\dagger_n$ are fermionic annihilation/creation operators satisfying 
$\left\{c_m,c^\dagger_n\right\}=\delta_{mn}$. The link operators $S^+_n$ are spin-raising operators. The parameters $t$ and $m$ control the hopping and mass respectively. We have introduced quadratic fermionic interactions governed by an interaction parameter $U$. The size of the system is denoted by $L$ and we will always consider periodic boundary conditions. 
In this model, there are no pure gauge terms as there are no plaquette terms in $1+1$ dimension and the electric field term is trivial for spin-$1/2$ ($E^2_n=\left(S^z_n\right)^2=1/4$).

We have a set of local symmetries as the Hamiltonian commutes with
$G_n=S^z_n-S^z_{n-1}-c^\dagger_nc_n$,
for all values of $n$. The Hilbert space breaks down into different sectors labeled by eigenvalues of $G_n$.
We will focus on the physical sector, where no external charges are present. Let $\rho_n= c^\dagger_nc_n+\left(\left(-1\right)^n-1\right)/2$ be the charge density. The physical sector is given by the states satisfying 
\begin{equation}
    \label{eq:gauss}
    \left(S^z_n-S^z_{n-1}\right)\left|\psi\right>=\rho_n\left|\psi\right>,
\end{equation}
where even sites can host zero or positive charge, and odd sites can host zero or negative charge.

This model exhibits $CP$-symmetry. Concretely, change conjugation ($C$)  is characterized by the fermionic transformation of the fields as $c_n\rightarrow \left(-1\right)^{n+1}c^\dagger_{n+1}$, $c^\dagger_n\rightarrow \left(-1\right)^{n+1}c_{n+1}$, $E_n\rightarrow -E_{n+1}$ and $U_n\rightarrow U^\dagger_{n+1}$. With the expressions properly understood under periodic boundary conditions, we can write the parity symmetry according to $c_n\rightarrow c_{-n}$, $c^\dagger_n\rightarrow c^\dagger_{-n}$, $E_n\rightarrow -E_{-n-1}$ and $U_n\rightarrow U^\dagger_{-n-1}$. The symmetries can be composed to give rise to the $CP$-symmetry, which we will study in Sec. \ref{sec:results}.

We end this section by noting that even though we focus on periodic boundary conditions, the algorithm can be straightforwardly generalized to open boundary conditions.

\section{The Meron-Cluster Approach}
\label{MCalgo}

Detailed expositions of the meron-cluster approach can be found in e.g \cite{chandrasekharan2003meron,ewertz2003loop}. Here we give a brief account on how and why it works. The derivation of the algorithm proceeds first through Trotter decomposition, followed by the decomposition of the weights into break-ups.

We generically describe expectation values of observables at a fixed inverse temperature $\beta$ as 
$\left<O\right>=\frac{1}{Z}\mathrm{Tr}\left(e^{-\beta H}\right)$ where $Z=e^{-\beta H}$.
In our problem, we set $H=H_1+H_2$ with these operators defined according to
\begin{align}
    H=H_1+H_2,\quad
    H_{1/2} = 
    \sum_{\mathrm{even/odd}}\left[
    -t \left(c^\dagger_n c_{n+1}+\mathrm{h.c.}\right)
    +U \left(c^\dagger_nc_n-\frac{1}{2}\right)
    \left(c^\dagger_{n+1}c_{n+1}-\frac{1}{2}\right)
    \right]. 
    \label{eq:meron-hamiltonian}
\end{align} 
Gauge fields and the mass term are not contemplated in the standard meron-cluster algorithm.
By partitioning $e^{-\beta H}=\left(e^{-\varepsilon H}\right)^N$, with $\varepsilon N=\beta$, we can approximate $e^{-\varepsilon H}\simeq e^{-\varepsilon H_1}e^{-\varepsilon H_2}$. By choosing a basis and introducing a resolution of the identity between any two pairs of exponentials, we build a $1+1$ path integral. Each configuration is characterized by a sequence of basis states ${\left\{\left|\psi_i\right>\right\}}_{i=0}^{2N}$, with $\left|\psi_0\right>\equiv\left|\psi_{2N}\right>$, which will receive a weight that is a product of the expectation values $\left<\psi_{i+1}\right|e^{-\varepsilon H_{1+\left(i\ \mathrm{mod}\ 2d\right)}}\left|\psi_{i}\right>$. We will generically denote this weight as $W\left(c_f\right)$, where $c_f$ condenses the collective dependence on all the $2N$ (fermionic) basis states of the configuration.

We adopt an occupation basis where the states are described by $\left|n_0\dots n_{L-1}\right>$, and $n_i\in\left\{0,1\right\}$ indicates the fermion occupation number at site $i$. $W\left(c_f\right)$ can be decomposed into a product of elementary plaquettes, i.e. elementary weights: $w_p\left(n,n^\prime;n^{\prime\prime}n^{\prime\prime\prime}\right)=\left<nn^\prime\right|e^{-\varepsilon H_i}\left|n^{\prime\prime}n^{\prime\prime\prime}\right>\equiv w_p\left(c\right)$. 

These weights are further decomposed in the form $w_p\left(c\right)=\sum_g w\left(c,g\right)$.    
In our case $g$ takes only two values, which we will denote by $v$ and $h$, referring to vertical and horizontal break-ups. 
We can fix this decomposition by demanding that all weights are non-negative and that $w\left(c,g\right)=w\left(c^\prime,g\right)\equiv w \left(g\right)$, where $c^\prime$ is obtained from $c$ by changing the occupation of any pair of vertical or horizontal neighbors whenever $g=v$ or $g=h$ respectively. These conditions imply $U=2t$. 
The decomposition of the weights of plaquettes induces a decomposition on the full configuration $W^\prime\left(c_f, G\right)$. All sites that are connected by break-ups form a cluster. By construction, clusters are represented by non-intercepting loops, and, when the occupation number of all elements of a cluster is changed at the same time, we obtain a configuration with the same weight. We call this process \emph{cluster flip}. We can now construct an algorithm that provides non-local updates of the configurations 
with high acceptance probability. 
The cluster algorithm follows two main steps: it updates the break-up decomposition and flips each resulting cluster with probability $p = 1/2$.
Meron-cluster algorithms require extra steps to solve the sign problem. 
These additional steps are no longer necessary if we want to sample only configurations satisfying Gauss' law, since this immediately eliminates all configurations with negative signs. 
In the next section, we will specify what conditions must be met to satisfy the Gauss' law.

\section{Constraints on Fermionic Configurations}
\label{sec:constraints_fermionic}

Every state of the gauge theory can be written as 
$\left|\psi\right>=\left|\psi_f\right>\otimes\left|\psi_s\right>$,
where $\left|\psi_f\right>$ encodes the fermionic state and $\left|\psi_s\right>$ the gauge field state (encoded by spins). We can then construct a path-integral using the same strategy described in the previous section, through Trotterization. For any fermionic configuration ${\left\{\left|\psi_{fi}\right>\right\}}_{i=0}^{2N}$, in the purely fermionic theory of the previous section, we can associate spin configurations ${\left\{\left|\psi_{si}\right>\right\}}_{i=0}^{2N}$ such that 
\begin{enumerate}
    \item $\left|\psi_{fi}\right>\otimes\left|\psi_{si}\right>$ satisfy Gauss' law;
    \item $\left<\psi_{fi+1}\right|\otimes\left<\psi_{si+1}\right|
    e^{-\varepsilon H_{1+\left(i\ \mathrm{mod}\ 2\right)}}
    \left|\psi_{fi}\right>\otimes\left|\psi_{si}\right>$ is non-zero for every $i$.
\end{enumerate}
For every fermionic configuration $c_f$, we can define a set of spin configurations ${\cal S}\left(c_f\right)$ that satisfy these conditions. Note that this does not exclude the possibility that ${\cal S}\left(c_f\right)$ is the empty set.

Let $W_{QLM}\left(c_f,c_g\right)$ be the weights associated with the gauge theory. By defining

\begin{equation}
    W_{eff}\left(c_f\right)=\sum_{c_g\in{\cal S}\left(c_f\right)}
    W_{QLM}\left(c_f,c_g\right),
    \label{eq:weff}
\end{equation}
we can reweight the standard cluster algorithm according to $W_{eff}\left(c_f\right)/W\left(c_f\right)$. The problem with this direct approach is that most configurations violate Gauss' law.



Suppose we inspect the link at $m$ and progress up through the lattice. From Gauss' law \eqref{eq:gauss}, if that spin is down ($S^z_{m}\rightarrow-1/2$), we will continue to encounter spin down as long as the subsequent charges are zero $\rho_{m^\prime}=0$. Once we encounter $\rho_n=1$, for some subsequent site $n$, the spin flips. We can never find $\rho_n=-1$ if at $n$ the spin is down as we would not be able to satisfy Gauss' law. For spin up, an analogous discussion takes place.
This implies that positive ($\rho_n=1$) and negative ($\rho_n=-1$) charges must alternate, including across the periodic boundary. Furthermore, for any valid sequence of positive and negative charges, the value of the gauge fields is automatically determined, except for the case where $\rho_n = 0$ for all $n$. We call this the \emph{reference state} $\left|\psi_R\right>$. Only configurations where all states correspond to the reference state will not uniquely determine the gauge fields. We call this the \emph{reference configuration} $c_R$, which admits two possible sets of values for the gauge fields.
Following this discussion, we can re-write \eqref{eq:weff} as 

\begin{equation}
    \label{eq:weff2}
    W_{eff}\left(c_{f}\right)=\left\{ 
    \begin{array}{ll}
        2W\left(c_{R}\right),\quad &c_{f}=c_{R}\\
        W\left(c_{f}\right),\quad &c_{f}\in{\cal V}\backslash\left\{ c_{R}\right\} \\
        0,\quad&\mathrm{otherwise}
    \end{array}
\right.,
\end{equation}
where ${\cal V}$ is the set of valid configurations, which we will now determine.


In principle, any algorithm valid for interacting fermions \eqref{eq:meron-hamiltonian} can be modified using \eqref{eq:weff2}. This translates to counting the reference configuration with weight factor 2, counting the remaining valid configurations with weight factor 1, and disregarding all the other configurations. Due to the large number of invalid configurations, this does not work in practice. In the next section, we will map the above-derived conditions for fermionic configurations into constraints on cluster flips. This will allow for the construction of a cluster algorithm where only valid combinations of cluster flips are allowed, and equivalently, only valid configurations $c\in {\cal V}$ are generated.

\section{Constraints on Cluster Flips}
\label{sec:constraints_flips}

\begin{figure}
    \centering
\includegraphics[width=.6\linewidth]{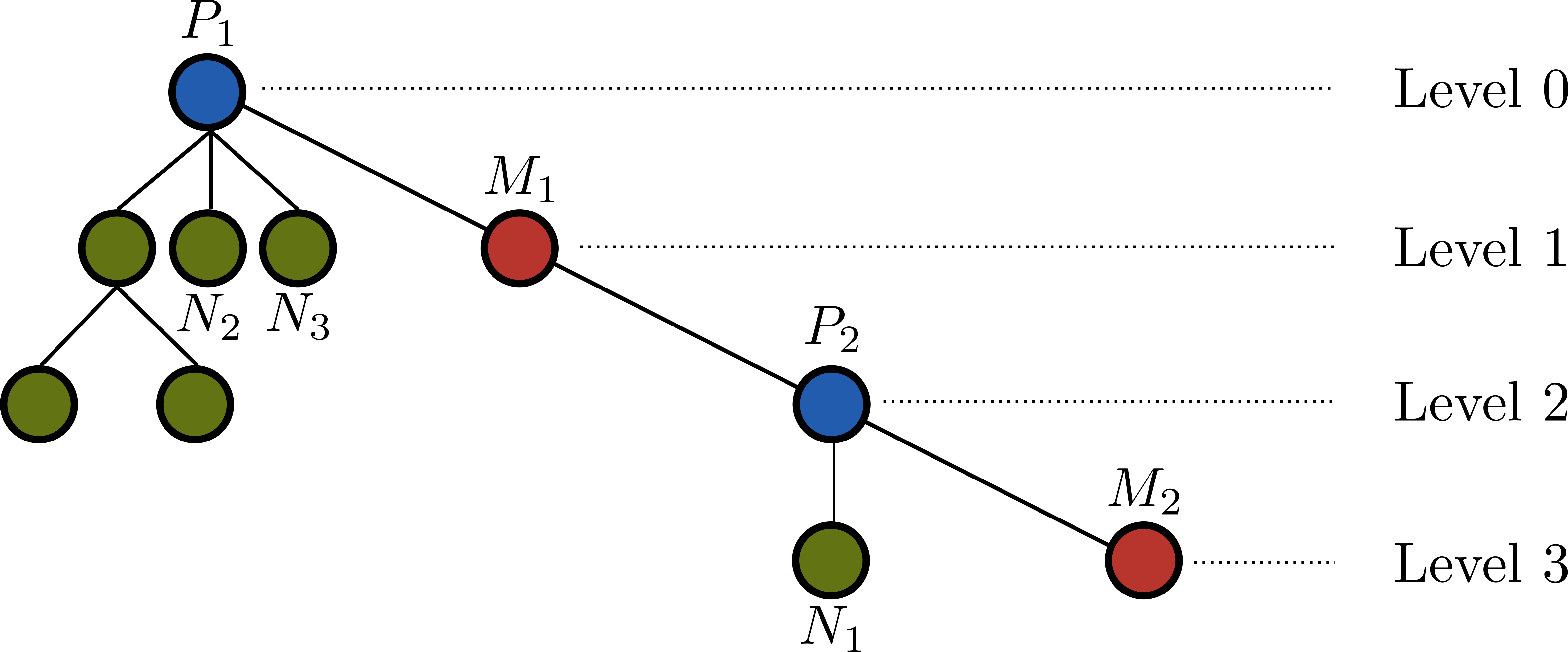}
\hspace{1cm}\includegraphics[width=0.23\linewidth]{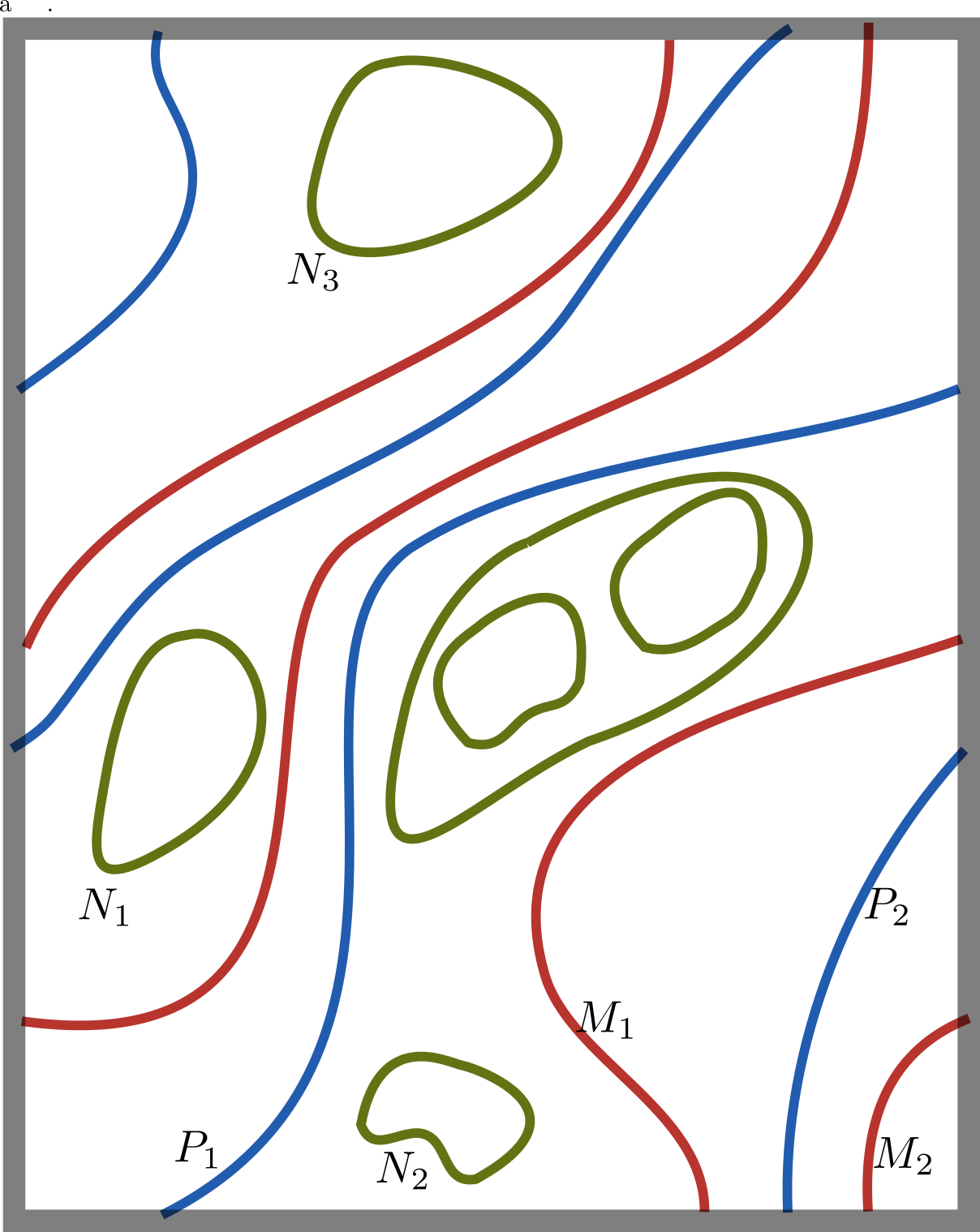}
    \caption{Example construction of a cluster tree with two positive clusters, $P_1$ and $P_2$, two negative clusters, $M_1$ and $M_2$, and six neutral clusters organized according to the rules of the text.}
    \label{fig:cluster_tree}
\end{figure}

To understand the constraints on cluster flips, that have to be satisfied to produce valid configurations, we need to recall some cluster properties. First, every site belongs to a single cluster and every cluster forms a  non-intersecting loop. Due to periodic boundary conditions in time and space, our space-time forms a torus.  We can then have winding cluster loops, which can wind around one or two directions of the torus, and non-winding loops. 
We can construct a hierarchy that describes how the loops are geometrically related. Clusters that wind in time (space) divide the torus into distinct areas. We take one winding cluster as the root of a tree. The other winding cluster that borders the area that is in the positive time (space) direction is a child of this cluster. The other cluster that borders the next area in the positive direction is then that cluster's child and so on until the first cluster is reached again, which orders all winding clusters. We call the $i$'th cluster we encounter in this procedure cluster $i$.  Non-winding clusters that lie in an area between winding clusters $i$ and $i+1$ are children of the cluster $i$.
If a non-winding cluster is nested in another, the nested cluster is a child of the cluster that surrounds it. If no charged clusters exist, the area outside all the clusters is taken as the root. An example of such a hierarchy can be found in Fig.~\ref{fig:cluster_tree}.

The second important property of clusters is that, if a site with an even spatial coordinate is occupied, then all sites with even spatial coordinates are occupied and all odd sites will be empty (and vice-versa). Therefore, either all sites are neutral (i.e. in the reference configuration) or all of them are charged. We will call clusters that are in the reference configuration unflipped and clusters charged on all sites flipped.

A flipped cluster introduces a loop of sites with charge $\pm 1$. All the links in either neighboring area of the cluster must therefore differ by $\pm 1$ according to Gauss' law. Due to the geometry of loops on a square lattice, and the fact that the charge of a site corresponds to its parity, the sign of the difference depends on the parity of the level of the cluster in the cluster tree (see Fig. \ref{fig:cluster_tree}).

This property leads to the following rule regarding valid cluster flips.
On a path from the root to a leaf, consecutive flipped clusters must be at odd distances. 
Additionally, the flipped clusters at the lowest levels need to have an even distance from one another, or equivalently, their levels need to have the same parity.
By construction, walking along a path on the torus corresponds exactly to a path on the tree, since neighboring clusters are connected.
Due to this property, when walking along any path, the links are now increased and decreased alternatingly while crossing the levels, which ensures only two link values are necessary.

 In the next section, we briefly describe an algorithm that samples these configurations with the correct distribution.

\section{The Constrained Cluster Algorithm}
\label{sec:constrained_algo}

As we have shown, sampling configurations of the gauge theory correspond to sample valid fermionic configurations uniformly (except for the reference configuration). This does not include the mass term of Eq. \eqref{eq:H}, which can however be introduced through reweighting.

We will make our decisions as we go from the root to the leaves. Once we get to cluster $i$, we know the value of the electric field in the region outside, which we will call $s$. We flip it with probability 

\begin{equation}
    p_i\left(s\right)=
    \frac
    {n_i\left(1,s\right)}
    {n_i\left(0,s\right)
    +n_i\left(1,s\right)}
    \label{eq:pflip}
\end{equation}
where $n_i\left(\sigma,s\right)$ is the number of valid ways of having $s$ in the outer region of cluster $i$, when $i$ is flipped ($\sigma=1$) or not ($\sigma=0$). These functions constitute a total of $4 n_c$ values. Once we have them, we can start from the root and work our way down the tree.

It remains to show that these values can be efficiently computed. Starting at the bottom of the tree, we can count how many ways that cluster can either be flipped or not given the electric field outside. This is possible because there is no further structure inside the cluster that could impact the value of the electric field. From there we can work our way up the tree, constructing systematically $n_i\left(\sigma,s\right)$ for every cluster. Once all these functions are obtained, we can start at the root of the tree and choose cluster orientations according to \eqref{eq:pflip}. As mentioned, these probabilities can be properly reweighted to account for non-zero mass, as cluster flips have a definite effect on the extra weight, independent of the other clusters. Any cluster, in/out of the reference configuration, receives a reweighting factor of $\exp\left(\pm\varepsilon m \left|{\cal C}\right|/4\right)$. With this in hand, we can study the ground-state properties of this model at varying mass.

\section{Numerical Results}
\label{sec:results}

\begin{figure}
    \centering
    \includegraphics[width=0.49\linewidth]{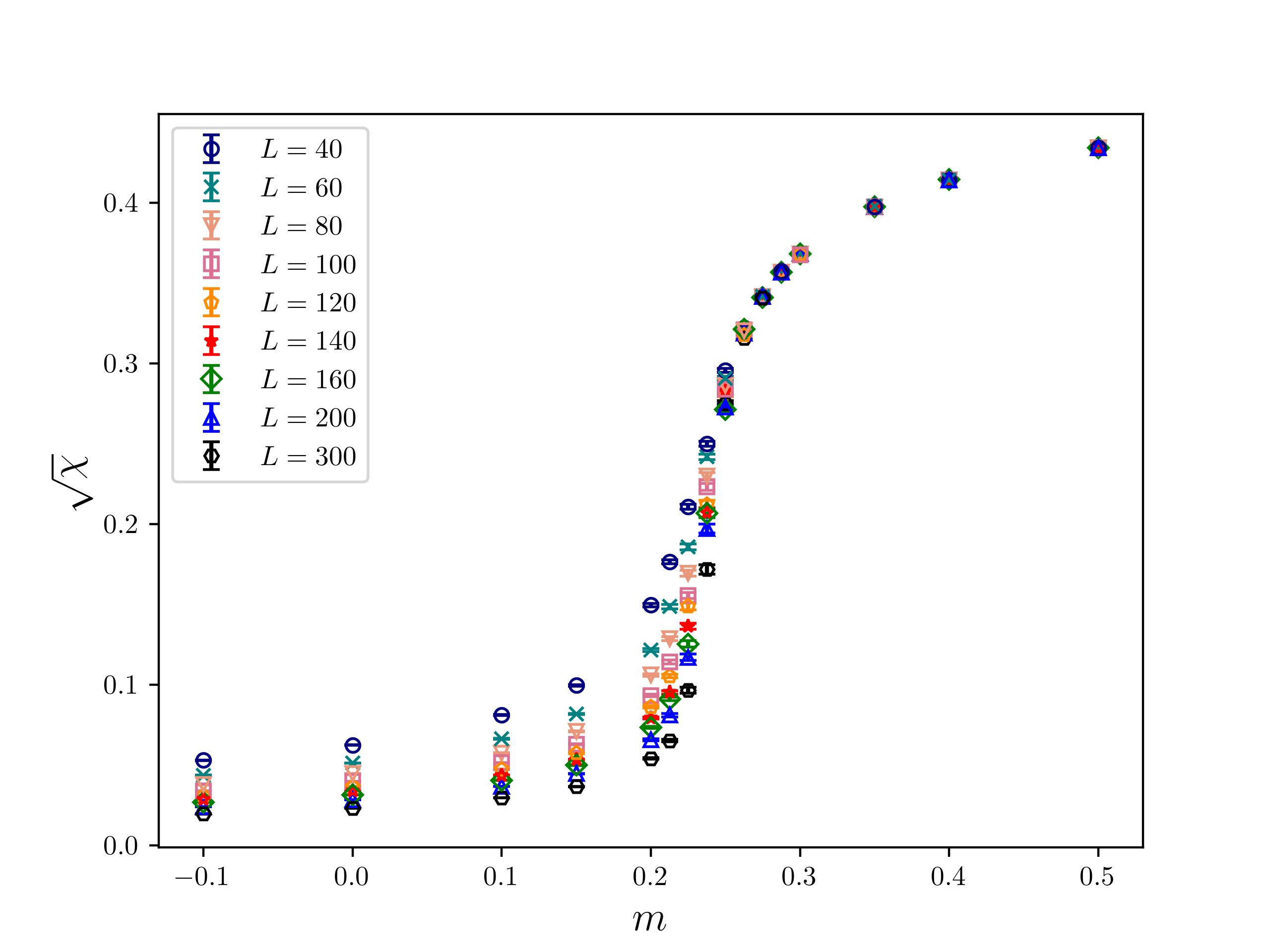}
    \includegraphics[width=0.49\linewidth]{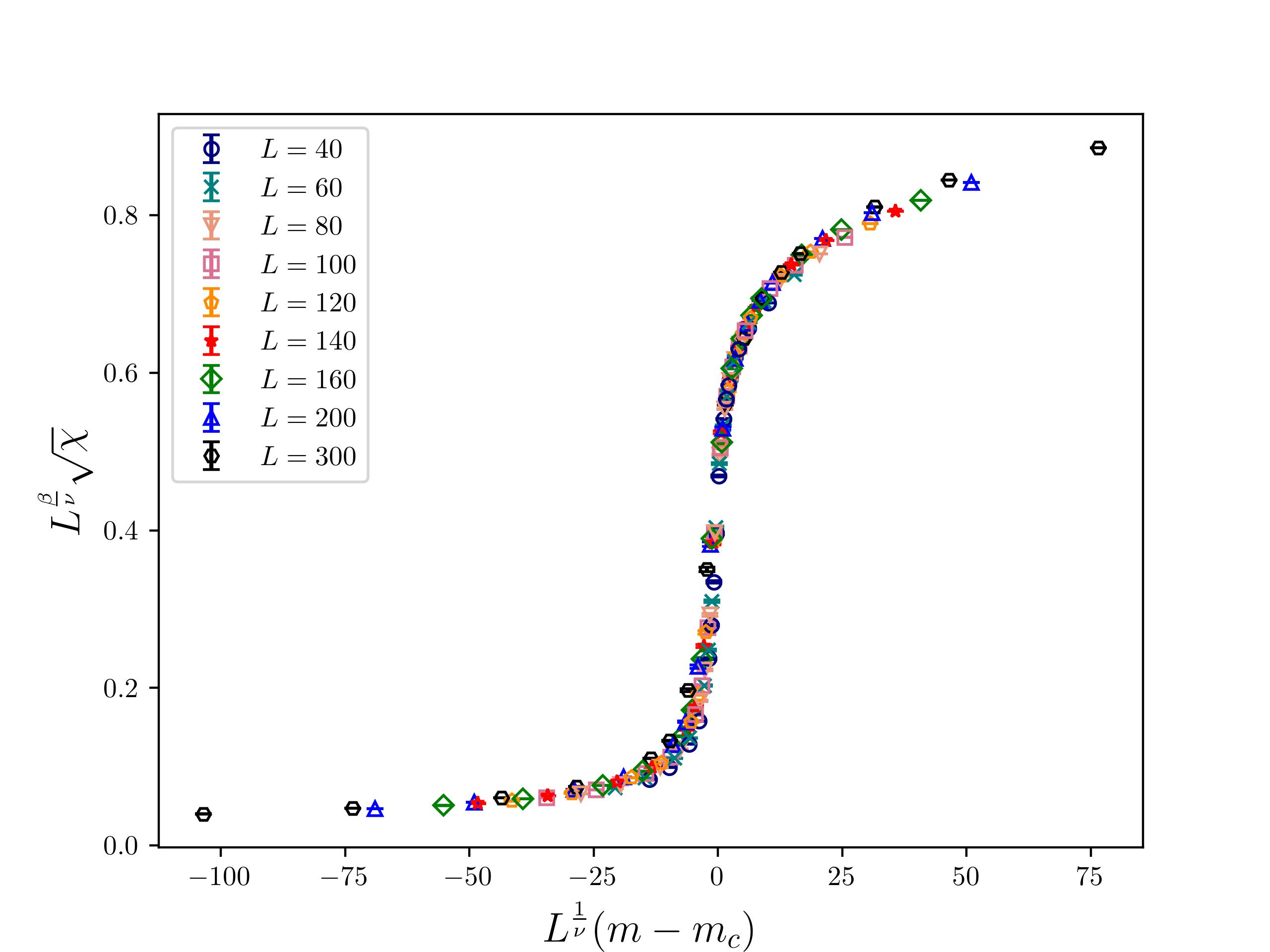}
    \caption{Left: Square root of the susceptibility for varying mass and lattice size. Right: Finite-size scaling of $\sqrt{\chi}_E$. The results are compatible with the Ising universality class, and a critical mass $m_c=0.245(3)$.}
    \label{fig:curve_collapse}
\end{figure}

In order to test the algorithm, we study the phase diagram of the model at varying mass. As discussed in Sec. \ref{sec:QLM} this model is $CP$-symmetric, which is a $\mathbb{Z}_2$ symmetry. For small values of the mass, pair creation is a low-energy process, and we expect the system to be in the symmetric phase, where there is no preferential direction of the electric field. This situation is characterized by a single ground state. In turn, when the mass is large enough, we expect the ground state to acquire a dominant direction of the electric field as charges scarcely appear. In this limit, the ground state is double degenerate, where the total electric flux acquires a definite value. The electric field susceptibility can be used as an order parameter. Concretely, we define
$
\chi_E=\left<\left(\sum_n E_n\right)^2\right>/L,
$
and plot its value for different volumes in Fig. \ref{fig:curve_collapse}. We observe evidence of a quantum phase transition compatible with the Ising universality class. From finite-size scaling analysis, we estimate the critical mass to be $m_c=0.245(3)$ and present the curve collapse for Ising critical exponents $\nu=1$ and $\beta=1/8$. This is in agreement with previous results using tensor networks \cite{rico2014tensor}. While the critical mass that we have found is different from the above-cited work, this is to be expected as this is a non-universal quantity and our model includes a fixed quartic fermionic interaction.

\section{Conclusion}
\label{sec:conclusion}

We have constructed a constrained cluster algorithm that is able to simulate fermionic systems coupled to gauge fields. It applies to spin-$1/2$ $U\left(1\right)$ quantum link model in $1+1$ dimensions. 
To simulate this model, we efficiently generate configurations of the pure fermionic theory with a cluster algorithm, while avoiding \emph{all} configurations incompatible with Gauss' law.

Our results allow for cross-validations with other methods, such as tensor networks, of theories directly relevant to quantum simulations of gauge theories, and are also useful for their benchmarking. These developments go hand-in-hand with providing an alternative route for the classical simulation of gauge theories. Further progress in this direction requires generalization to larger spins, and other gauge groups. Another interesting question is whether these ideas can suitably be applied in higher dimensions. This would be a challenging task since there the gauge field becomes truly dynamical.

\section*{Acknowledgements}
We are grateful to D. Banerjee, S. Chandrasekharan, E. Huffman and U.-J. Wiese for inspiring discussions. We acknowledge access to Piz Daint at the Swiss National Supercomputing Centre, Switzerland under the ETHZ’s share with the project ID eth8. Support from the Google Research Scholar Award in Quantum Computing and the Quantum Center at ETH Zurich is gratefully acknowledged. This work was performed at the Aspen Center for Physics, which is supported by National Science Foundation grant PHY-2210452.

\bibliographystyle{JHEP}
\bibliography{bibliography}

\end{document}